\documentclass[prl,twocolumn,preprintnumbers,amsmath,amssymb]{revtex4}
\usepackage{graphicx}
\usepackage{xspace}
\newcommand{\degC}{\ensuremath{^{\circ}\,\textrm{C}}\xspace}
\newcommand{\degK}{\ensuremath{^{\circ}\,\textrm{K}}\xspace}
\usepackage{dcolumn} %for  two-column only
\begin{document}

\title{Step-edge instability during epitaxial growth of graphene from SiC(0001)}

%\date{\today}

\author{Valery Borovikov}
\email{valery.borovikov@physics.gatech.edu}
\author{Andrew Zangwill}
\email{andrew.zangwill@physics.gatech.edu}
\affiliation{School of Physics \\ Georgia Institute of Technology,  Atlanta, Georgia 30332, USA}

\maketitle

{\bf The unique electronic properties of graphene offer the possibility that it could replace silicon when microelectronics evolves to nanoelectronics.\cite{Castro} Graphene grown epitaxially on silicon carbide\cite{Hass}  is particularly attractive in this regard because SiC is itself a useful semiconductor and, by suitable manipulation of the growth conditions,\cite{Hass06,Virojanadara,Emtsev09} epitaxial films can be produced that exhibit  all the transport properties of ideal, two-dimensional graphene desired for device applications.\cite{Orlita08,Miller09} Nevertheless, there is little or no understanding of the actual kinetics of growth, which is likely to be required for future process control.
As a step in this direction, we propose a local heat release mechanism to explain finger-like structures observed when graphene is grown  by step flow decomposition of SiC(0001). Using a continuum equation of motion for the shape evolution of a moving step, a linear stability analysis  predicts whether a shape perturbation of a straight moving step grows or decays as a function of growth temperature, the background pressure of Si maintained during growth, and the effectiveness of an inert buffer gas  to retard the escape of Si atoms from the crystal surface. The theory gives semi-quantitative agreement with experiment for the characteristic separation between fingers observed when graphene is grown in a low-pressure induction furnace or under ultrahigh vacuum conditions.}

\begin{figure}
\includegraphics[width=8.0cm, keepaspectratio] {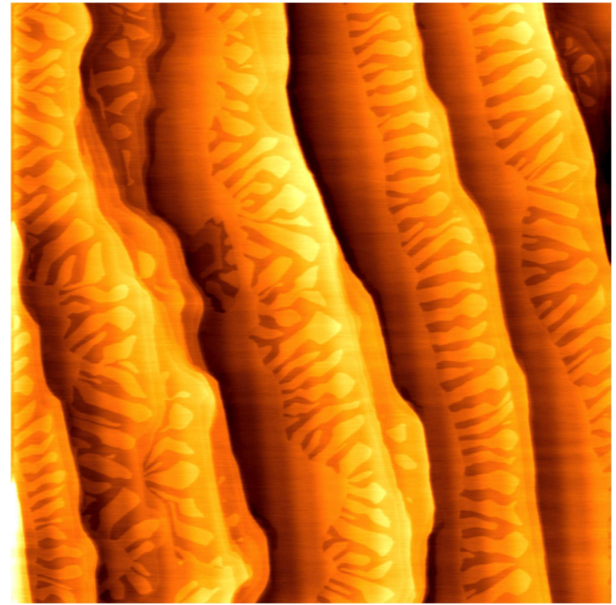}
\caption{\label{fig:Fig1}  AFM image from Ref.~\onlinecite{unpublished} of a SiC(0001) surface graphitized at $T  \sim 1600\degC$ in an induction furnace with a background Si pressure $P  \sim 1\times10^{-5}$ torr. The field of view is 10 $\mu$m. The steps descend from left to right.  }
\end{figure}

\begin{figure}
\begin{center}\includegraphics[
width=7.0cm, keepaspectratio] {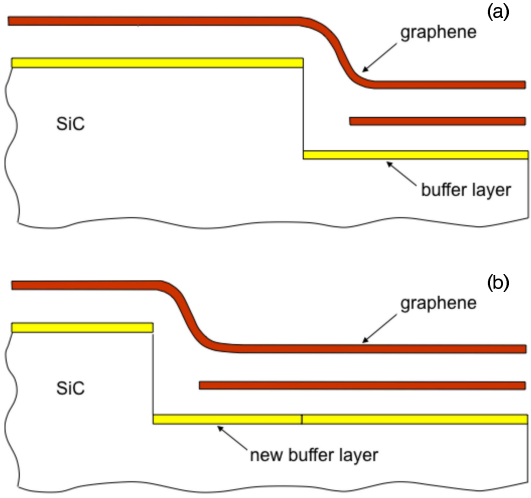}
\caption{\label{fig:Fig2}  {Graphene growth at a step on SiC(001): (a) initial state;   (b) final state.
}}
\end{center}
\end{figure}

Figure~\ref{fig:Fig1} is an atomic force microscope  image of 2-3 ML of graphene grown  on a commercial wafer of SiC(0001) at $\sim 1600\degC$ in a low-pressure induction furnace. The original  wafer was H-etched at high temperature to produce a regular array of steps.\cite{Feenstra2,Borovikov} When heated above the graphitization temperature, $T_G$, the steps bunch into  macrosteps (which run vertically in Figure~\ref{fig:Fig1}) separated by large nearly flat terraces. A continuous and conformal film of graphene covers the entire sample. A step with a highly ramified edge appears between each pair of adjacent macrosteps. A similarly ramified  step morphology has been reported by Hupalo and co-workers for graphene grown on  a vicinal surface of SiC(0001) under ultrahigh vacuum (UHV) conditions.\cite{Hupalo}
The thermal decomposition process that produces graphene occurs preferentially at steps, where Si and C atoms are least well bonded.  Figure~\ref{fig:Fig2}(a) shows a step on SiC(0001) covered conformally by one complete layer of graphene and a second incomplete layer.\cite{Lauffer}  The diagram reflects the fact that,  above $\sim 1100\degC$,  the bulk of the SiC crystal is ``buffered'' from anything that lies above (vacuum or graphene)  by a carbon-rich $6\sqrt{3}\times 6\sqrt{3}$ reconstruction.  The structure of the buffer layer is not completely known, but it apparently resembles graphene itself, albeit distorted so it bonds strongly to the silicon layer immediately below.\cite{Mattausch_PRL_07} When the temperature exceeds $1200\degC$, the SiC beneath the  buffer layer of the upper terrace begins to decompose. The liberated Si atoms desorb, the upper buffer layer transforms to graphene, and the liberated C atoms recrystallize at the base of the step to extend the lower buffer layer. Figure~\ref{fig:Fig2}(b) shows the advance of the second graphene layer.

\begin{figure}
\begin{center}\includegraphics[
width=8.0cm, keepaspectratio] {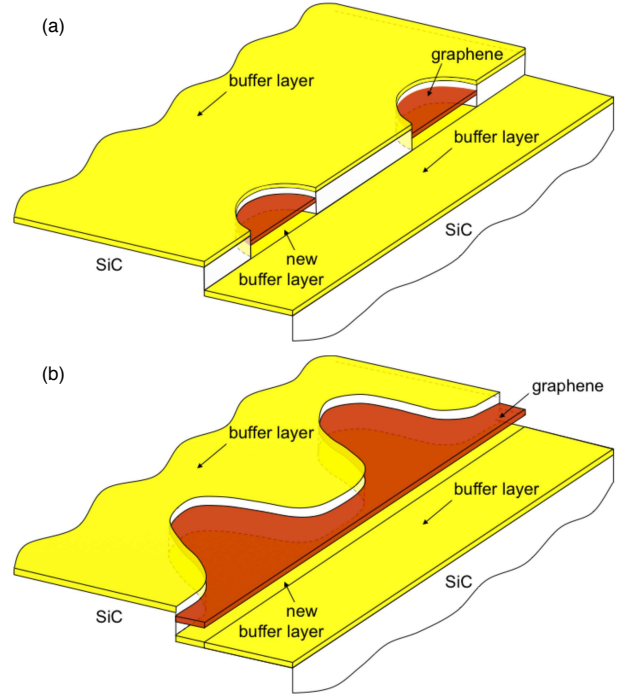}
\caption{\label{fig:Fig3}  {Onset of instability: (a)  thermal decomposition of SiC starts at random points along the  step-edges; (b) the formation of  new buffer layer releases heat, which locally accelerates further decomposition and graphitization. The topmost complete graphene layer is not shown.
}}
\end{center}
\end{figure}

It is important to appreciate that surface diffusion on the terraces plays very little  role in the growth scenario sketched in Figure~\ref{fig:Fig2}. Therefore, despite their appearance, the fingers  in Figure~\ref{fig:Fig1} cannot  arise from a Mullins-Sekerka type morphological instability driven by surface diffusion and an asymmetry in attachment/detachment kinetics at the steps.\cite{Bales,Pimpinelli} We suggest an alternative mechanism based on the fact that the crystallization of free carbon atoms into new buffer layer material  creates strong  $\sigma$-bonds. The accompanying release of heat should transiently increase the local temperature and induce further decomposition. This is a positive feedback mechanism that promotes further decomposition at points along the step where it has already begun.  Since graphene production accompanies decomposition, Figure~\ref{fig:Fig3} shows how the step edge between graphene and the upper buffer layer  may be expected to evolve as time goes on.  Comparison with Figure~\ref{fig:Fig1}  identifies the light-colored fingers as  buffer layer  and the darker channels between them as graphene.
To be more quantitative, we let $h(x,t)$ denote the position of the  SiC step edge and take account of the local heating effect (phenomenologically) by supposing that the decomposition rate depends on the local curvature of the front; a higher rate for concave regions and lower rate for convex regions. Therefore, if $V$ is the average velocity of the step due to decomposition, a suitable evolution equation for $h(x,t)$ is
\begin{equation}
\frac{\partial h}{\partial t}= -V - a V \frac{\partial^{2} h}{\partial x^2} + \sigma \Gamma  \frac{\partial^{2} h}{\partial x^2}  - \sigma D \frac{\partial^{4} h}{\partial x^4}. \label{StepMotionEquation}
\end{equation}
The last two terms in (\ref{StepMotionEquation}) are familiar from extensive studies of the effect of capillary smoothing on the morphology of step edges on vicinal surfaces.\cite{Jeong, Pimpinelli} The second derivative term models evaporation-condensation events where atoms detach from a step edge, migrate rapidly on the adjacent terrace, and re-attach to the step elsewhere. The fourth derivative term models edge diffusion events where atoms migrate along the step edge itself. In (\ref{StepMotionEquation}), $a$ is the SiC lattice constant, $\sigma=a^3 \gamma/kT$ ($\gamma$ is the SiC step stiffness), $\Gamma  = \nu \exp(-E_1/kT)$ is the mean rate at which atomic species detach from a straight SiC step  ($\nu$ is an attempt frequency) and $D=a^2\nu \exp(-E_2/kT)$ is the edge diffusion constant.

Kinetic theory relates the average step velocity to the difference
between the flux of Si atoms subliming from the surface (as measured by the equilibrium vapor pressure
of Si over SiC) and the background pressure $P$ in the growth chamber:
\begin{equation}
 V = \beta \frac{a^{3}}{\sqrt { 2 \pi m k_{B} T } } (P_{\rm eq} - P).
\label{stepVelocity}
\end{equation}
In this expression,  $m$ is the mass of a Si atom,  ${\rm log_{10}{\it P}_{eq}(Pa) = 12.74 - 2.66 \times 10^4/T (\degK)}$,\cite{Lilov} and $\beta$ is the evaporation coefficient. \cite{Hirth}
We note that increasing $P$ shifts $T_G$ to higher temperature and no decomposition occurs if $P>P_{\rm eq}$.\cite{Tromp}
\begin{figure}
\begin{center}\includegraphics[
width=6.2cm, keepaspectratio] {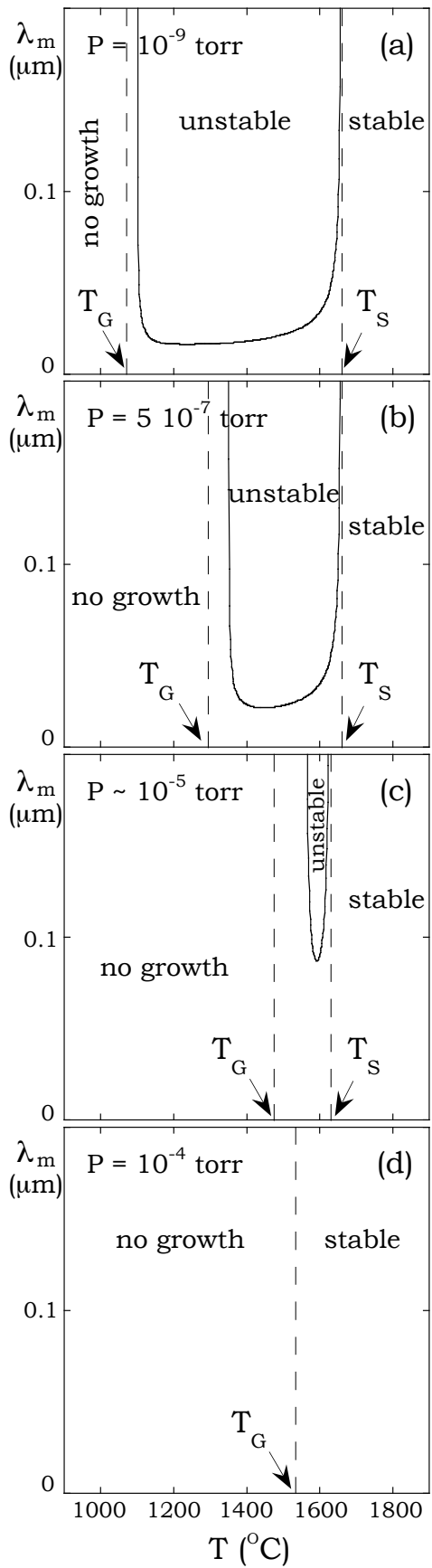}
\caption{\label{fig:Fig4}  {Dependence of the fastest growing wavelength $\lambda_{m}$ on the growth  temperature $T$ for several values of the background Si pressure $P$.
}}
\end{center}
\end{figure}
It is straightforward to perform a linear stability analysis of (\ref{StepMotionEquation}) assuming a solution of the form $h(x,t)=-Vt+\epsilon(t) \sin(2\pi x/\lambda)$. The perturbation amplitude $\epsilon(t)$ growth (decays) exponentially when the wavelength $\lambda$ is greater (less) than a critical wavelength. The most unstable (fasting growing) wavelength is
\begin{equation}
\lambda_m = \sqrt{\frac{8\pi^2 \sigma D }{aV - \sigma \Gamma }}. \label{CriticalWL}
\end{equation}
The constants that enter  $\lambda_m$ that require special discussion are $\gamma$,  $E_1$, $E_2$, and $\beta$, since none of these is truly known for SiC.  For the step stiffness of a three-bilayer step, we chose $\gamma = 1~{\rm  eV}/{\rm \AA}$, which is approximately three times the stiffness of steps on Si(111).\cite{Jeong} The detachment and surface diffusion barriers were chosen as $E_1=5.6~$eV and $E_2=5.0~$eV based on values for  surface diffusion barriers for C atoms and Si atoms on several SiC surfaces given by Fissel.\cite{Fissel} The evaporation coefficient was fixed at $\beta=0.046$ to make the calculated growth rate equal the experimental growth rate associated with Figure~\ref{fig:Fig1}.

Figure~\ref{fig:Fig4} is a plot of $\lambda_m$ as a function of growth temperature for several values of Si background pressure. There is no graphene growth when  $T<T_G$ and step edges are absolutely stable (no fingers) when $T>T_S$.  Between  $T_G$ and $T_S$, there is a temperature window where the step edge is unstable. This window is largest for UHV growth and decreases as $P$ increases. There is a critical pressure, $P_S$, above which graphene growth occurs with no morphological instability of the SiC step edges. Below $P_S$ and just above $T_G$,  there is a narrow region of stable growth. All these features arise from the competition between the destabilizing effect of local heating and the stabilizing effect of step edge evaporation/condensation and edge diffusion. The capillary forces of step smoothing always prevail when $T>T_S$. The narrow temperature range of  edge stability just above $T_G$  occurs because the destabilizing curvature term in (\ref{StepMotionEquation}) is proportional to $V$, which is very small near $T_G$.  Otherwise, local heating dominates and the scenario sketched in Figure~\ref{fig:Fig3} occurs.

For the growth conditions stated in Figure~\ref{fig:Fig1}, Figure~\ref{fig:Fig4}(c) predicts a most unstable wavelength of $\sim$ 0.1 $\mu$m. This agrees well with the separation between fingers in the AFM image. Additional experiments where graphene was grown in a furnace with a higher value of background pressure produced only straight steps,\cite{unpublished} in qualitative agreement with Figure~\ref{fig:Fig4}(d). Figure~\ref{fig:Fig4}(a) corresponds to an ultrahigh vacuum environment and so should be relevant to the graphene growth experiments reported in  Ref.~\onlinecite{Hupalo}. At  $T\approx 1200\degC$, step edge fingers were observed with a mean separation of  $\lambda_m \sim$ 0.01 $\mu$m, which accords well with our theory.  The fingers disappeared when the UHV experiments were repeated at $T \sim 1350\degC$,\cite{Miron} which agrees only semi-quantitatively with Figure~\ref{fig:Fig4}(a). On the other hand, the value of the evaporation coefficient  $\beta$ in (\ref{stepVelocity}) is sample-dependent and probably differs between  the induction furnace experiment and the UHV experiment.\cite{caveat}  A reduction of $\beta$ by only a factor of two is sufficient to close the window in Figure~\ref{fig:Fig4}(a) enough to match the UHV data.
\begin{figure}
\begin{center}\includegraphics[
width=7.4cm, keepaspectratio] {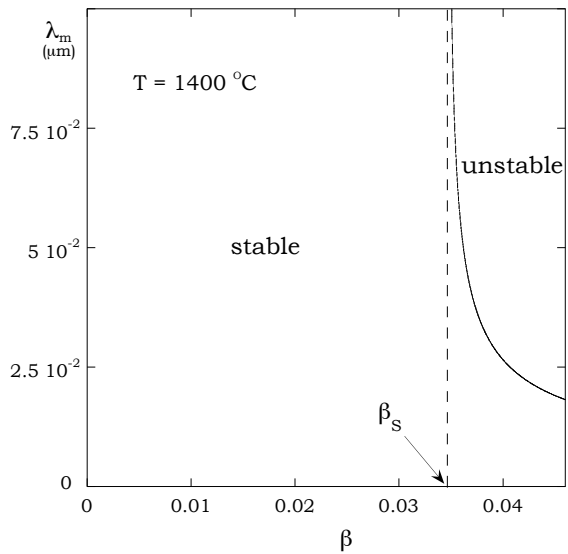}
\caption{\label{fig:Fig5}  {Dependence of the fastest growing wavelength $\lambda_{m}$ on the evaporation coefficient $\beta$ for $P=0$ and  $T = 1400~^{\circ}$C.
}}
\end{center}
\end{figure}
Another process variable used in recent graphene growth experiments  is the pressure of an inert gas intentionally introduced into the growth chamber.\cite{Virojanadara,Emtsev09} The presence of such a gas was found to tremendously improve the quality of the graphene produced epitaxially on SiC(0001). Presumably, the effect of the inert gas is to retard the escape of Si atoms from the decomposing crystal. The same thing happens in the present  model if, as above,  we reduce the  value of the evaporation coefficient $\beta$ in (\ref{stepVelocity}). Accordingly,  Figure~\ref{fig:Fig5} shows the dependence of the most unstable wavelength on $\beta$ at $T=1400\degC$ and $P=0$.  There is a critical value,  $\beta_{S}$, below which the theory predicts graphene growth without a step edge instability. We encourage experiments to test this prediction.

In summary, we have shown that finger-like step edges observed during the growth of epitaxial graphene by step flow decomposition of  SiC(0001) can be understood from a competition between capillary smoothing and a curvature-driven mechanism for step edge roughening. We suggest (but cannot prove) that the origin of the roughening is accelerated step decomposition by local heating at points where it has  already begun. The heat source is  the formation energy of a graphene-like buffer layer that accompanies the growth of graphene on this surface. A linear stability analysis of a step equation of motion that takes account of all these effects predicts the separation between fingers as a function of the growth process variables: temperature, background Si pressure, and the influence of an inert gas in the growth chamber. The theory agrees semi-quantitatively with growth experiments conducted in both a low-pressure induction furnace and under ultrahigh vacuum conditions.

\begin{acknowledgments}
The authors are grateful to M. Sprinkle and W.A. de Heer for permission to show the experimental data in Figure~\ref{fig:Fig1}. They also acknowledge useful discussions with Phillip First, Edward Conrad, Miron Hupalo, Fan Ming, and Ming Ruan. The work of V.B. was supported by the Department of Energy under Grant No. DE-FG02-04-ER46170 and the MRSEC program of the National Science Foundation under Grant No. 0820382.
\end{acknowledgments}


\begin{thebibliography}   {99}

\bibitem{Castro} Neto, A. H. C., Guinea, F., Peres, N., Novoselov, K., and Geim, A. The electronic properties of graphene. \textit{Rev. Mod. Phys.} {\bf 81,} 109\---162 (2009).

\bibitem{Hass} Hass, J., de Heer, W. A., and Conrad, E. H. The growth and morphology of epitaxial multilayer graphene. \textit{J. Phys. Cond. Matter} {\bf 20,} 323202 (2008).

\bibitem{Hass06} Hass, J. \textit{et al.} Highly ordered graphene for two dimensional electronics. \textit{Appl. Phys. Lett.} {\bf 89,} 143106 (2006).

\bibitem{Virojanadara} Virojanadara, C. \textit{et al.} Homogeneous large-area graphene layer growth on 6H-SiC(0001). \textit{Phys. Rev. B} {\bf 78,} 245403 (2008).

\bibitem{Emtsev09} Emtsev, K. V. \textit{et al.} Towards wafer-size graphene layers by atmospheric pressure graphitization of silicon carbide. \textit{Nature Mater.} {\bf 8,} 203\---207 (2009).

\bibitem{Orlita08} Orlita, M. \textit{et al.} Approaching the dirac point in high-mobility multilayer epitaxial graphene. \textit{Phys. Rev. Lett.} {\bf 101,} 267601 (2008).

\bibitem{Miller09} Miller, D. L. \textit{et al.} Observing the quantization of zero mass carriers in graphene. \textit{Science} {\bf 324,} 924\---927 (2009).

\bibitem{unpublished} Sprinkle, M. and de Heer, W. A. (unpublished).

\bibitem{Feenstra2} Nie, S. \textit{et al.} Step formation on hydrogen-etched 6H-SiC\{0001\} surfaces. \textit{Surf. Sci.} {\bf 602,} 2936\---2942 (2008).

\bibitem{Borovikov} Borovikov, V. and Zangwill, A. Step bunching of vicinal 6H-SiC\{0001\} surfaces. \textit{Phys. Rev. B} {\bf 79,} 245413 (2009).

\bibitem{Hupalo} Hupalo, M., Conrad, E., and Tringides, M. C. Growth mechanism for epitaxial graphene on vicinal 6H-SiC(0001) surfaces. \textit{Phys. Rev. B} {\bf 80,} 041401 (2009).

\bibitem{Lauffer} Lauffer, P. \textit{et al.} Atomic and electronic structure of few-layer graphene on SiC(0001) studied with scanning tunneling microscopy and spectroscopy. \textit{Phys. Rev. B} {\bf 77,} 155426 (2008).

\bibitem{Mattausch_PRL_07} Mattausch, A. and Pankratov, O. Ab initio study of graphene on SiC. \textit{Phys. Rev. Lett.} {\bf 99,} 076802 (2007).

\bibitem{Bales} Bales, G. S. and Zangwill, A. Morphological instability of a terrace edge during step-flow growth. \textit{Phys. Rev. B} {\bf 41,} 5500 (1990).

\bibitem{Pimpinelli} Pimpinelli, A. and Villain, J. \textit{Physics of Crystal Growth}. (Cambridge University Press, New York, 1998).

\bibitem{Jeong} Jeong, H.-C., Williams, E. D. Steps on surfaces: experiment and theory. \textit{Surf. Sci. Rep.} {\bf 34,} 171\---294 (1999).

\bibitem{Lilov} Lilov, S. K. Thermodynamic analysis of phase transformations at the dissociative evaporation of silicon carbide polytypes. \textit{Diamond Relat. Mater.} {\bf 4,} 1331\---1334 (1995).

\bibitem{Hirth} Hirth, J. P. and Pound, G. M. Coefficients of evaporation and condensation. \textit{J. Phys. Chem.} {\bf 64,} 619\---626 (1960).

\bibitem{Tromp} Tromp, R. M. and Hannon, J. B. Thermodynamics and kinetics of graphene growth on SiC(0001). \textit{Phys. Rev. Lett.} {\bf 102,} 106104 (2009).

\bibitem {Fissel} Fissel, A. Artificially layered heteropolytypic structures based on SiC polytypes: molecular beam epitaxy, characterization and properties. \textit{Phys. Rep.} {\bf 379,} 149\---255 (2003).

\bibitem{Miron} Hupalo, M. (private communication).

\bibitem{caveat} The kinetic pathway by which liberated Si atoms escape into the vapor from the sub-surface decomposition region is unknown and likely depends on the details of the sample preparation and history.

\end{thebibliography}
\end{document}